\documentclass[aps,prc,preprint,groupedaddress]{revtex4-1}

\usepackage{graphicx} 
\usepackage{amsmath}
\usepackage{braket}
\usepackage{bm}

\newcommand{\Oxy}{${}^{16}{\rm O}$}
\newcommand{\Sal}{${}^{32}{\rm S}$}
\newcommand{\Si}{${}^{28}{\rm Si}$}
\newcommand{\OO}{${}^{16}{\rm O}$+${}^{16}{\rm O}$ }
\newcommand{\aSi}{$\alpha$+${}^{28}{\rm Si}$ }

\begin{document}


\title{$\bm \alpha$ + ${}^{\bf 28}{\bf Si}$ and ${}^{\bf 16}{\bf O}$+${}^{\bf 16}{\bf O}$
molecular states,\\ and their isoscalar monopole strengths}

\author{Masaaki Kimura}
\email{masaaki@nucl.sci.hokudai.ac.jp}
\affiliation{Department of Physics, Hokkaido University, Sapporo 060-0810,
Japan} 
\affiliation{Nuclear Reaction Data Centre (JCPRG), Hokkaido University, Sapporo
060-0810, Japan} 
\affiliation{Research Center for Nuclear Physics (RCNP), Osaka University, Ibaraki
567-0047, Japan} 
\author{Yasutaka Taniguchi}
\email{taniguchi-y@di.kagawa-nct.ac.jp}
\affiliation{Department of Information Engineering, National Institute of Technology
(KOSEN), Kagawa College, Mitoyo 769-1192, Japan} 
\affiliation{Research Center for Nuclear Physics (RCNP), Osaka University, Ibaraki
567-0047, Japan} 
\date{\today}

\begin{abstract}
 The properties of the \aSi and \OO molecular states which are embedded in the excited
 states of \Sal~and can have an impact on the stellar reactions, are investigated using
 the antisymmetrized molecular dynamics. From the analysis of the cluster
 spectroscopic factors, the candidates of  \aSi and \OO molecular states are
 identified close to  and above the cluster threshold energies. The calculated
 properties of the \aSi molecular states are consistent with those reported by the \aSi
 resonant scattering  experiments. On the other hand, the  \OO molecular
 state, which is predicted to be identical to the superdeformation of \Sal, is
 inconsistent with the assignment proposed by an $\alpha$ inelastic scattering
 experiment. Our calculation suggests that the monopole transition from the
 ground state to the \OO molecular state is rather weak and is  not strongly excited
 by the $\alpha$  inelastic scattering.  
\end{abstract}

\maketitle

\section{Introduction}
The \aSi and \OO molecular
states~\cite{Baye1976,Ando1980,Baye1984,Langanke1984,Kondo1989} which is embedded in the  
excited states of \Sal~are fascinating subjects in nuclear cluster physics and nuclear
astrophysics. The $\alpha$-induced reactions such as
${}^{28}{\rm Si}(\alpha,\gamma)^{32}{\rm S}$ and
${}^{28}{\rm Si}(\alpha,p)^{31}{\rm P}$ play an important role in the silicon burning
process of the stellar evolution and nucleosynthesis~\cite{Bodansky1968}. The \aSi
molecular states, if they exist at the incident energy, increase the reaction rate in order
of magnitude and determine the reaction
products~\cite{Smulders1964,Toevs1971,Rogers1977}. In a similar manner, the \OO molecular
states crucially affect the oxygen burning
process~\cite{Spinka1974,Kovar1979,Hulke1980,Wu1984,Thomas1986,Weaver1993,ElEid2004,
Diaz-Torres2007,Gasques2007}. Furthermore, the \OO molecular states have  unique and
interesting characteristics from the view point of nuclear cluster physics. Many
theoretical studies~\cite{Freer1995,Ohkubo2002,Kimura2004,Maruhn2006,Ichikawa2011,
Ebran2014,Ray2016,Ebran2017} predicted that an \OO molecular band should exist
just below the \OO threshold energy, and it must be identical to the superdeformed state
of \Sal. Although the superdeformation of \Sal~has not been observed, the theoretical
prediction sheds a new light on the clustering of light nuclei. 

Experimentally, these molecular states have been explored using the ordinary techniques
such as transfer 
reactions~\cite{Maher1972,Lindgren1974,Peng1978,Berg1979,Tanabe1981,Morita1985} and 
resonant scattering~\cite{Gai1981,Manngard1994,Kallman1994,Kallman2003,Lonnroth2010}.
However, as the molecular states are embedded in the continuum of \Sal, it is difficult
to identify them from many other resonances. This  difficulty prevents us from the
full understanding of the molecular states and the superdeformation. 

In this decade, instead of the ordinary experimental techniques, the isoscalar
monopole and dipole transitions induced by $\alpha$ inelastic scattering are attracting
a lot of research interest to overcome the above-mentioned problem. 
These transitions can populate the deep sub-barrier resonances and have unique
selectivity for molecular states; hence, they are effective to identify the molecular
states embedded in the
continuum~\cite{Kawabata2007,Kanada-Enyo2007,Yamada2008,Chiba2016}. In particular, the
method has already been successfully applied to the discussion of clustering and
molecular states in many stable and unstable
nuclei~\cite{Funaki2008,Ito2011,Yamada2012,Ichikawa2012,Yang2014,Kanada-EnYo2014,
Yamada2015,Chiba2015,Chiba2017b,Nakao2018,Kanada-EnYo2019,Chiba2020,Kanada-Enyo2020,
Baba2020}. On the same line of physics, Itoh {\it et al.}~\cite{Itoh2013} have measured
the isoscalar transitions of $^{32}{\rm S}$, identified several excited states
with enhanced transition strengths, and proposed a new band assignment for the \aSi and
\OO molecular states (and hence the superdeformed states of \Sal). 

In this work, motivated by the new and interesting experimental data, we theoretically
investigated the \aSi and \OO molecular states, and their monopole strengths.
The framework of the antisymmetrized molecular dynamics
(AMD)~\cite{Kanada-Enyo2003,Kanada-Enyo2012,Kimura2016} has already been applied to the
study of the molecular states and superdeformation of $sd$-$pf$
nuclei~\cite{Kimura2004,Kanada-Enyo2005,Kimura2006,Taniguchi2007,Taniguchi2009}. Recently 
it has been extended to handle the rotation  effect of the deformed clusters and
successfully applied to investigate the ${}^{12}{\rm C} + {}^{16}{\rm O}$ molecular
states at deep sub-barrier energy~\cite{Taniguchi2020}. Following these studies, we
extended our researches considering both the \aSi and \OO channels in addition to the
rotation effect of the deformed \Si~cluster. It was found that the monopole transition
has a strong selectivity to  \aSi molecular states, but it is insensitive to \OO 
molecular states. Consequently, we conclude that many of the excited states reported
by Itoh {\it et al.}~\cite{Itoh2013} should be attributed to the \aSi molecular
states. From the systematics of the cluster  spectroscopic factors and $B(E2)$
transition strengths, we also propose the assignment of the \aSi and \OO molecular bands. 

This paper is organized as follows: In the next section, we explain the AMD framework
and how we handle both the \aSi and \OO channels, as well as the rotation effect of the
deformed \Si~cluster. In section~\ref{sec:result}, we explain the \aSi and \OO wave
functions obtained by the variational calculations. We discuss the properties of the
$0^+$ states and their monopole strengths in comparison with the observations. We also
suggest the \aSi and \OO molecular band assignment. The final section summarizes this
work.

\section{Theoretical framework}\label{sec:framework}
The theoretical framework used in this paper is the same as our previous work for
the   $^{12}{\rm C}$+$^{16}{\rm O}$ molecular states. The deformed-basis AMD is combined
with the $d$-constraint method. The Hamiltonian is expressed as,  
\begin{align}
 H = \sum_{i=1}^A t(i) + \sum_{i<j}^A v_{NN}(ij) + \sum_{i<j}^A v_C(ij)  - t_{\rm c.m.}, 
\end{align}
where the Gogny D1S parameter set~\cite{Berger1991} is used for the effective
nucleon-nucleon interaction $v_{NN}$ and the Coulomb interaction $v_{C}$ is approximated
by a sum of seven Gaussians. The center-of-mass kinetic energy $t_{\rm c.m.}$ is properly
removed from the Hamiltonian without any approximation.  This Hamiltonian reasonably
describes the threshold energies of interest without any adjustment. The \aSi and \OO
threshold energies measured from the \Sal~ground state are calculated as 7.56 and 16.25
MeV, respectively, which are compared with the experimental data 6.95 and 16.54 MeV.    

The variational wave function of the deformed-basis AMD is a parity-projected Slater
determinant of the single-particle wave packets~\cite{Kimura2004a}, 
\begin{align}
 \Phi &= \mathcal{A}\set{\varphi_1,...,\varphi_A},\\
 \varphi_i=&\prod_{\sigma=x,y,z}\left(\frac{2\nu_\sigma}{\pi}\right)^{1/4}\exp
 \set{-\nu_\sigma\left(r_\sigma - Z_{i\sigma}\right)^2}\nonumber\\
 &\times  \left(\alpha_i\ket{\xi_\uparrow}+\beta_i\ket{\xi_\downarrow}\right)
 \times \left(\ket{p}\ \text{or}\ \ket{n}\right),
\end{align}
where each of the $\varphi_i$ has the deformed Gaussian form and has the parameters: the 
centroid of the Gaussian $\bm{Z}_i$, size parameter $\bm \nu$ and spin direction
$\alpha_i$ and $\beta_i$.  The isospin part is fixed to either of proton  or
neutron. The size parameter $\bm \nu$ is a real-valued vector, but the other parameters
are complex-valued. They are determined by the energy variation with two different
constraints. The first one is the constraint on the quadrupole deformation parameter
$\beta$, which we call the $\beta$-constraint. It is noted that the $\beta$-constraint
was already used to study the superdeformation of \Sal~and the \OO molecular states
within the AMD framework~\cite{Kimura2004}. The second constraint is imposed on the 
inter-cluster distance between the $\alpha$ and ${}^{28}{\rm Si}$ clusters, and between
two $^{16}{\rm O}$ clusters, which we call the $d$-constraint~\cite{Taniguchi2004}. We
classify the centroids of the wave packets into two groups corresponding to the cluster
configurations, and define an approximate inter-cluster distance $d$ as the distance
between the  center-of-masses of two groups. For example, $d$ for \aSi configuration
is defined as,
\begin{align}
 d = \left|\frac{1}{4}\sum_{i\in \alpha}Re\bm Z_i 
 - \frac{1}{28}\sum_{i\in ^{28}{\rm Si}}Re\bm Z_i\right|.
\end{align}
The value of $d$ is constrained from 2 fm to 8 fm with the interval of 0.5 fm. It is
emphasized that the $d$-constraint is essential for describing the \aSi molecular states,
because the $\beta$-constraint yields only the mean-field and  \OO molecular
states. Furthermore, it can handle the cluster polarization effect and
the rotation effect of the deformed clusters in a natural manner. 

From the energy variation with the constraints, we obtain the wave functions which have
the minimum energies for each given value of $\beta$ or $d$. After the energy variation,
the wave functions are projected to the eigenstates of the angular momentum, and
superposed to diagonalize the  Hamiltonian (generator coordinate method; GCM). 
\begin{align}
 \Psi^{J\pi}_{M\alpha}= \sum_{iK} b_{iK\alpha}P^{J}_{MK}\Phi^\pi(\beta_i) +
 \sum_{iK} d_{iK\alpha}P^{J}_{MK}\Phi^\pi(d_i),\label{eq:gcmwf}
\end{align}
where $P^{J}_{MK}$ denotes the angular momentum projector, $\Phi^\pi(\beta_i)$ and
$\Phi^\pi(d_i)$  are the wave functions obtained by the $\beta$- and $d$-constraints,
respectively.  The coefficients of the superposition $b_{iK}$ and $d_{iK}$ are
determined by solving the Hill-Wheeler equation~\cite{Hill1953}.

As a measure for the \aSi and \OO clustering, we calculate the reduced width amplitude
(RWA), which is the probability amplitude to find the clusters at the 
inter-cluster distance $a$. It is defined as the overlap between the reference cluster
wave function and the GCM wave function given by Eq. (\ref{eq:gcmwf}),
\begin{align}
y_\ell(a) = \sqrt{\frac{32!}{C_1! C_2!}}
 \Braket{\frac{\delta(r-a)}{r^2}[\Phi_{C_1}\Phi_{C_2}Y_{\ell}(\hat r)]^J_M|
  \Psi^{J\pi}_{M\alpha}},\label{eq:rwa1}
\end{align}
where $C_{1,2}$ and  $\Phi_{C_1,C_2}$ denote the masses and wave functions of clusters,
respectively: $C_1=4$ and $C_2=28$ for the \aSi configurations, and $C_1=C_2=16$ for the
\OO configurations. The reference wave function (bra state) describes the 
system in which two clusters with the masses of $C_1$ and $C_2$ are mutually orbiting
with the angular momentum $\ell$ and the inter-cluster distance $a$. Here, the wave
functions of the $\alpha$ and \Oxy~clusters are assumed to be the 
the harmonic oscillator wave functions with the doubly closed shell structure, which
reproduce the observed charge radii. The wave function of the \Si~cluster is
approximated using a single AMD wave function projected to either of  $J^\pi=0^+$ or $2^+$
states. The Eq. (\ref{eq:rwa1}) was calculated using the Laplace expansion
method~\cite{Chiba2017} for the \aSi channel and using the projection method to the
Brink wave function~\cite{Horiuchi1972,Horiuchi1977} for the \OO channel. The
cluster spectroscopic factors in these channels are calculated by the
squared integral of $y_{\ell}$.  
\begin{align}
 S_{\ell} = \int_0^\infty da\ a^2 |y_{\ell}(a)|^2,
\end{align}
which is enhanced for the developed cluster states, and is used to identify
the molecular states. 

In this work, we focus on the isoscalar monopole ($IS0$) transition strength which has
been regarded and utilized as a novel probe for the molecular states in
stable and unstable nuclei. The transition operator is defined as follows,
\begin{align}
 \mathcal{M}^{IS0} = \sum^A_{i=1} r'^{2}_i.
 \end{align}
Note that the single-particle coordinate $\bm r'_i$ is measured from the center-of-mass
$\bm r_{\rm c.m.}$, {\it i.e.}  $\bm r'_i\equiv \bm r_i - \bm r_{\rm c.m.}$; hence, our
calculation is free from the spurious center-of-mass contributions. The reduced transition
matrix from the ground state to the excited $0^+$ state is calculated as
\begin{align}
 M(IS0;0^+_1\rightarrow 0^+_{\rm ex.})=
 \braket{\Psi(0^+_{\rm ex.})|\mathcal{M}^{IS0}|\Psi(0^+_{\rm g.s.})} ,
\end{align}
where $\Psi(0^+_{\rm g.s.})$ and  $\Psi(0^+_{\rm ex.})$ are the wave functions of the
ground and excited $0^+$ states, respectively.

\section{Results and discussions}\label{sec:result}
\subsection{Molecular configurations obtained by the variational calculations}
\begin{figure}[tbp]
\includegraphics[width=0.6\hsize]{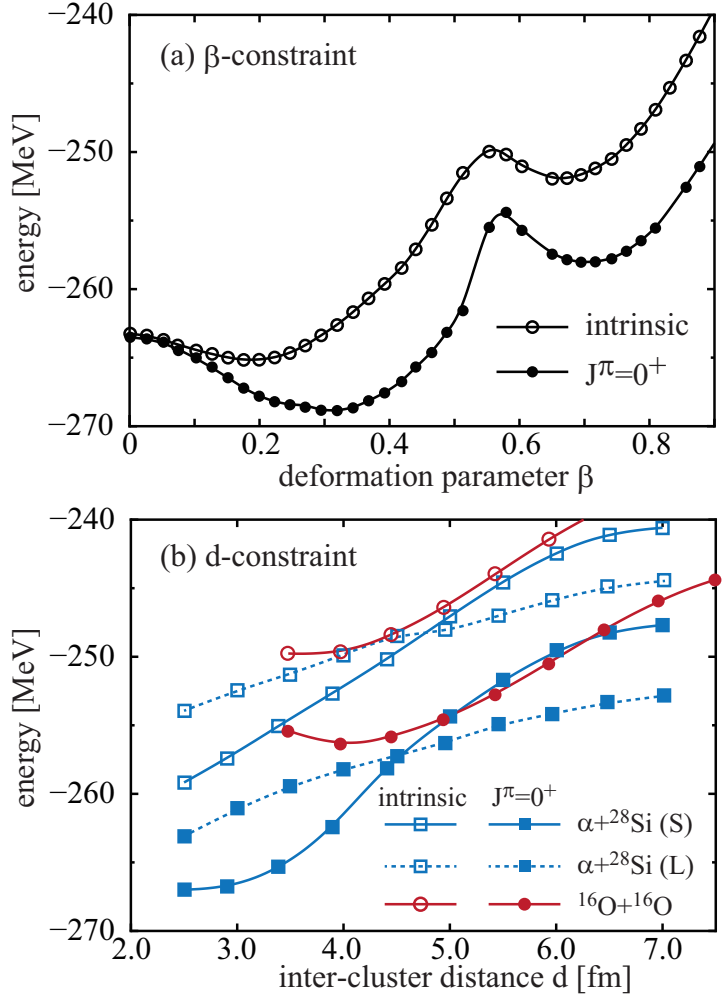}
 \caption{(color online)  Energy curves of the positive-parity states before (intrinsic)
 and after the  angular  momentum projection to $J^\pi=0^+$.  (a) The energy curves
 obtained by the  constraint on the quadrupole deformation parameter $\beta$.  (b) The
 energy curves of  the \aSi and \OO molecular configurations.} 
 \label{fig:surface}
\end{figure}
\begin{figure}[tbp]
\includegraphics[width=0.6\hsize]{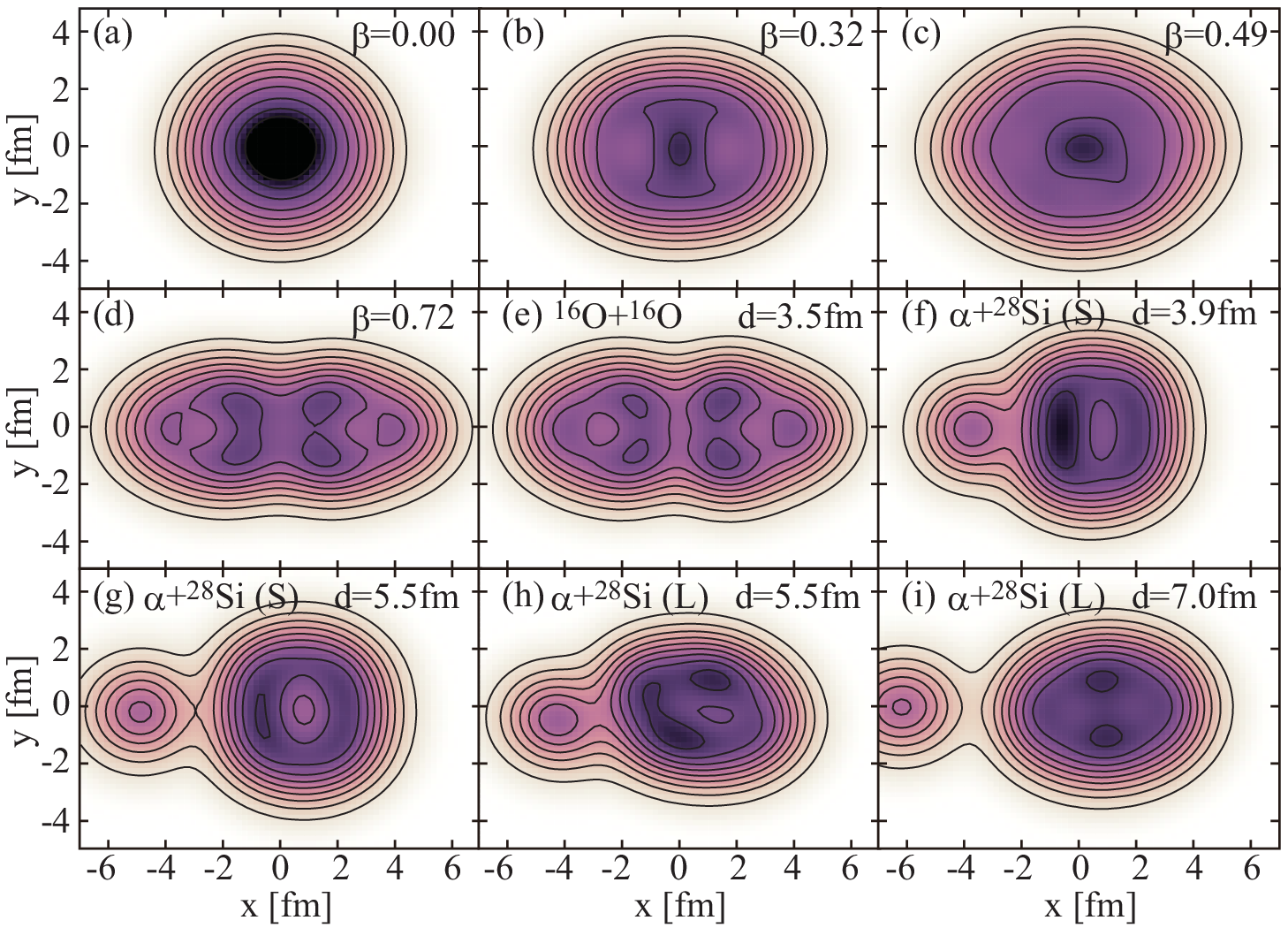}
 \caption{(color online) (a)-(d):  The intrinsic densities obtained by the
 $\beta$-constraint. The  panels (b) and  (d) correspond to  the ground and
 superdeformed minima on the  energy surface, respectively. (e)-(i): The intrinsic
 densities obtained by the  $d$-constraint. The panel (e)  shows the intrinsic  density
 of the \OO configuration at  the energy minimum. The panels (f) and (g) are the S-type
 \aSi configurations in which the symmetry axis of the \Si~cluster is  parallel to the
 $x$-axis, while the panels (h) and  (i) are the L-type configurations in which  the
 shortest axis is perpendicular to the $x$-axis.}
 \label{fig:density}
\end{figure}

Figure~\ref{fig:surface} (a) shows the energy curves of the positive-parity states
obtained by the $\beta$-constraint. It has two minima at $\beta=0.32$ and 0.72 after the
angular momentum projection, which correspond to the ground state and superdeformed
state, respectively. Their intrinsic density distributions shown in
Fig.~\ref{fig:density} (b) and (d) appear considerably different and impress the exotic
shape of the superdeformed minimum  (panel (d)) which is extremely deformed with a neck
and two-centered. Similar density distributions of the superdeformed state have also
been reported by many other theoretical 
studies~\cite{Rodriguez-Guzman2000,Inakura2002,Bender2003,Kimura2004,Maruhn2006,
Ichikawa2011,Ebran2014,Ray2016,Ebran2017}. 

The energy curves for the \aSi and \OO molecular configurations obtained by the
$d$-constraint are shown in Fig.~\ref{fig:surface} (b). We have obtained two different
\aSi molecular configurations which have different orientations of the deformed
\Si~cluster. We call them S- and L-type configurations in the following. In the S-type
configuration  denoted by (S), the symmetry axis of the 
oblate deformed \Si~ cluster is parallel to the $x$-axis on which the center-of-mass of
$\alpha$ and \Si~clusters are placed (see Fig.~\ref{fig:density} (f) and (g)). On the
contrary, in the L-type configuration (L), the symmetry axis of \Si~is perpendicular to
the $x$-axis (see Fig.~\ref{fig:density} (h) and (i)). By mixing both the S- and L-type
configurations, we can handle the rotation effect of the deformed \Si~cluster 
within the AMD framework.  It is  also noted that these two configurations have different 
single-particle structure at a small inter-cluster distance. The S-type configuration
approaches the ground state configuration ($0\hbar\omega$) at a short inter-cluster
distance; hence, its energy ($E=-267.0$ MeV at $d=2.5$ fm) is close to that of the
ground state minimum  ($E=-268.9$ MeV at $\beta=0.25$) as shown in
Fig.~\ref{fig:surface}. It is important to note that the squared overlaps of the wave
functions between the ground state and the S-type \aSi configurations are non-negligible
after the parity and  angular-momentum projection to $J^\pi=0^+$, although they appear
quite different at a glance. Indeed, the overlap between the wave functions shown in
Figs.~\ref{fig:density} (b) and (f) is as large as 0.45. On the contrary, the L-type
configuration approaches  a $4\hbar\omega$ excited configuration at a small distance.
Consequently, the L-type configurations are orthogonal to the ground state configuration
and their  energies are relatively higher than the S-type configurations. It is  noted
that these different asymptotics of the molecular configurations play a crucial role
for the isoscalar monopole transitions.

The \OO configuration is almost identical to the superdeformed state (a
$4\hbar\omega$ configuration) at the energy minimum ($d=3.5$ fm), and its energy
($-257.6$ MeV) is very close to that of the superdeformed minimum ($-258.0$ MeV). It is
impressive that their density distributions are significantly similar to each other
(Fig.~\ref{fig:density} (d) and (e)), and the squared overlap of their wave functions is
as large as 0.92 which indicates that they are actually identical. This is the reason why
many theoretical 
studies~\cite{Rodriguez-Guzman2000,Ohkubo2002,Kimura2004,Maruhn2006,Ichikawa2011,Ebran2014,
Ray2016,Ebran2017}
discuss the similarity of the superdeformation of \Sal~and the \OO molecular
states. However, despite the consistent and convincing discussions by many
theories,  experimental information about the superdeformation of \Sal~had been rather
limited~\cite{Morita1985,Curtis1996}. Recently, Itoh {\it et al.}~\cite{Itoh2013}
provided a new report by investigating the isoscalar monopole transition strengths
of \Sal. In particular, based on the observed strong monopole transitions, they proposed
a new assignment of the \OO molecular states, and hence the superdeformed states. We will
verify their assignment in the following sections.  

\subsection{Molecular states and their monopole strengths}

\begin{figure}[tbp]
\includegraphics[width=0.6\hsize]{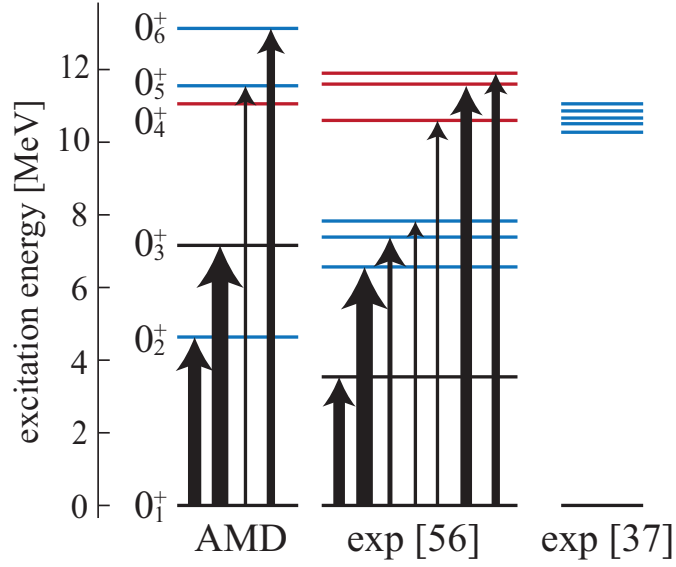}
 \caption{(color online) The calculated and observed~\cite{Lonnroth2010,Itoh2013}
 candidates of the  \aSi (blue lines) and \OO (red lines) molecular states with
 $J^\pi=0^+$. The widths of  the arrows are  proportional to the isoscalar monopole
 transition matrix.}  
 \label{fig:zero}
\end{figure}

\begin{table}[htb]
  \caption{The calculated excitation energies in MeV, isoscalar monopole transition
 matrices in $\rm fm^2$ and cluster spectroscopic factors of the $0^+$
 states. $S_{\alpha\ell=0}$,  $S_{\alpha\ell=2}$ and $S_{\rm O}$ denote the
 spectroscopic factors in the  $\alpha$+${}^{28}{\rm Si}(0^+_1)$, $\alpha$+${}^{28}{\rm
 Si}(2^+_1)$ and  ${}^{16}{\rm O}$+${}^{16}{\rm O}$ channels, respectively. 
 The observed excitation energies and the isoscalar monopole matrices~\cite{Itoh2013}
 are also listed.}  \label{tab:mono}  
 \begin{ruledtabular} 
  \begin{tabular}{cccccccc}
   \multicolumn{6}{c}{AMD}& \multicolumn{2}{c}{exp. }\\
   \cline{1-6}\cline{7-8}
   &$E_x$ &$M(IS0)$ & $S_{\alpha,\ell=0}$ & $S_{\alpha,\ell=2}$ & $S_{\rm O}$ & $E_x$ &
   $M(IS0)$\\\hline 
   $0^+_1$ & 0.0  &      & 0.09 & 0.04  & 0.00  & 0.0   &    \\
   $0^+_2$ & 4.6  & 5.7  & 0.05 & 0.06  & 0.00  & 3.78  & 4.0\\
   $0^+_3$ & 7.0  & 6.5  & 0.02 & 0.01  & 0.02  & 6.59  & 6.3\\
           &      &      &  &   &       & 7.65  & 3.8\\
           &      &      &  &   &       & 7.95  & 2.7\\
   $0^+_4$ & 11.0 & 0.0  & 0.02 & 0.01  & 0.32  &       &    \\
   $0^+_5$ & 11.6 & 2.8  & 0.29 & 0.14  & 0.00  & 11.49 & 3.3\\
   $0^+_6$ & 13.1 & 4.8  & 0.34 & 0.12  & 0.02  & 11.62 & 5.4\\
          &      &      &      &       &       & 11.90 & 4.3\\
  \end{tabular}
 \end{ruledtabular}
\end{table}

\begin{figure}[tbp]
\includegraphics[width=0.6\hsize]{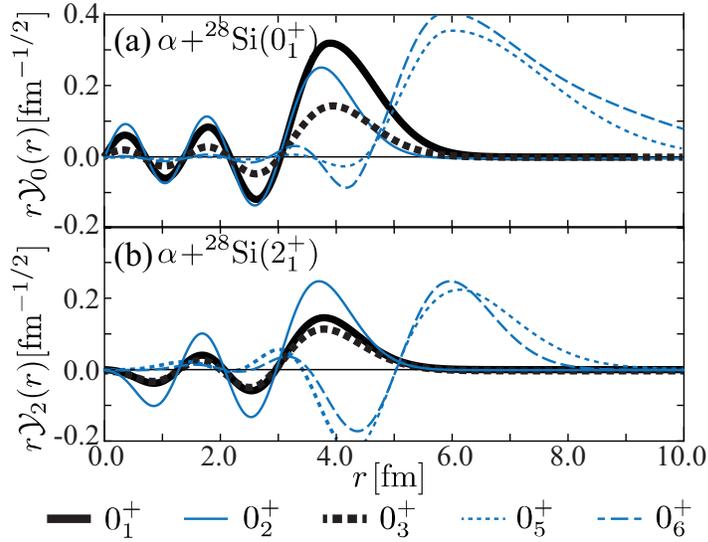}
 \caption{(color online) The reduced width amplitudes in the $\alpha$+${}^{28}{\rm Si}(0^+_1)$ and
 $\alpha$+${}^{28}{\rm Si}(2^+_1)$ channels calculated for the $0^+_1$, $0^+_2$,$0^+_3$,
 $0^+_5$ and $0^+_6$ states, respectively.} 
 \label{fig:rwa}
\end{figure}

In this section, we focus on the $J^\pi=0^+$ states, and discuss their molecular
structure, monopole  transition strengths and experimental candidates. However, before
the discussion of the present results, it may be useful to summarize the
experimental information about the \aSi and \OO molecular states. Many resonances which
are the candidates of the \aSi molecular states have been reported above the \aSi
threshold energy by the resonant scattering
experiments~\cite{Manngard1994,Kallman1994,Kallman2003,Lonnroth2010}. 
In particular, L\"onnroth {\it et al.}~\cite{Lonnroth2010}
comprehensively summarized the observed resonances covering broad energy region, and
proposed an \aSi molecular band. The candidates of the $0^+$ resonances they proposed
are fragmented into many states in between 10.25 and 11.05 MeV as shown in
Fig.~\ref{fig:zero}.  They all have the $\alpha$ decay widths ranging 
from a few keV to a few tens keV, and many of them coincide with the resonances observed in
other experiments~\cite{Manngard1994,Kallman1994,Kallman2003}. The $\alpha$
transfer reaction~\cite{Maher1972,Lindgren1974,Peng1978,Berg1979,Tanabe1981} is another
useful probe for the \aSi molecular states, especially for the states below the
decay threshold. Peng. {\it et al.}~\cite{Peng1978,Berg1979} and Tanabe
{\it et al.}~\cite{Tanabe1981} reported  the $\alpha$ spectroscopic factor of the
$0^+_2$ states by means of the $(^{16}{\rm O},{}^{12}{\rm C})$ and $(^{6}{\rm Li},d)$
reactions, respectively. They concluded that the $\alpha$ spectroscopic factor
of the $0^+_2$ state is approximately 0.50-0.75  relative to that of the ground
state (it varies, depending on the incident energy). Tanabe {\it et al.} also reported
that several states at 10 to 11 MeV are strongly populated by the $({}^{6}{\rm Li},d)$
reaction, and hence, suggested as the candidates of the  \aSi molecular states. Although
the spin-parity assignment was not discussed, it is important to note that the
energies of these states are very close to the $0^+$ resonances reported by L\"onnroth 
{\it et al.}~\cite{Lonnroth2010}.  

The isoscalar monopole strength is a novel probe for the molecular states, and has an
unique selectivity. Itoh {\it et al.}~\cite{Itoh2013} measured the isoscalar monopole
transitions of \Sal~by the $\alpha$ inelastic scattering, and 
reported several states as the candidates of the molecular states.  In addition to the 
$0^+_2$ state, they found that six excited states have 
the enhanced monopole strength as listed in Table~\ref{tab:mono}. They are classified
into two groups; three states at  6 to 8 MeV and the other three at 10 to 12
MeV. As summarized in Fig.~\ref{fig:zero}, the former group was proposed as the \aSi
molecular states, and the latter as the \OO molecular states. Furthermore, the latter
group, the states at 10 to 12 MeV, is also proposed as the superdeformed states, because
the \OO molecular state and the superdeformation should be identical. To summarize the
experimental data, the $0^+$ resonances at 10 to 12 MeV are observed in many
experiments. They are assigned as the \aSi molecular states in 
Refs.~\cite{Tanabe1981,Manngard1994,Kallman1994,Kallman2003,Lonnroth2010}, but are
assigned as the \OO molecular states in Ref.~\cite{Itoh2013}. Itoh {\it et al.} also
reported another group of the states at 6 to 8 MeV and assigned them as the \aSi
molecular states.

Now, we discuss the present numerical results in comparison with the above-mentioned
experimental data. The calculated ground state is predominated by the mean-field 
configuration shown in Fig.~\ref{fig:density} (b). The squared overlap between the ground
state and this configuration is 0.92. It is noted that the ground state
also has a large overlap with the S-type \aSi molecular configurations with  small
inter-cluster distances. The overlap between the ground state and the \aSi configuration
shown in Fig.~\ref{fig:density} (f) is as large as 0.46 and the calculated spectroscopic
factors of the ground state are $S_\alpha=0.09$ and 0.04 in the 
$\alpha$+${}^{28}{\rm Si}(0^+_1)$ and $\alpha$+${}^{28}{\rm Si}(2^+_1)$ channels,
respectively. This indicates that the $\alpha$ cluster correlation exists even in
the ground state. In fact, the calculated RWA of the ground state (Fig. \ref{fig:rwa})
has a peak at 3 to 4 fm showing the $\alpha$ cluster formation at the nuclear
surface. These results qualitatively agree with the observed large cross section of 
${}^{32}{\rm S}(p,p\alpha)^{28}{\rm Si}$~\cite{Carey1984} which is sensitive to
the $\alpha$ cluster formation at the surface of the ground
state~\cite{Roos1977,Yoshimura1998,Yoshida2016,Yoshida2018,Yoshida2019}. On the
other hand, the ground state has no overlap with the L-type \aSi and \OO configurations
as they asymptotically approach the $4\hbar\omega$ excited configurations at zero
inter-cluster distance, and are almost orthogonal to the ground state.

The $0^+_2$ state largely consists of almost the spherical configuration shown in
Fig.~\ref{fig:density} (a), and their squared overlap is 0.67. In addition, it also
has a non-negligible overlap with the S-type \aSi molecular configuration.  The overlap
between the $0^+_2$ state and the \aSi configuration shown in Fig.~\ref{fig:density} (f)
is 0.22, which indicates the non-negligible $\alpha$ cluster correlation in this
state. The calculated RWA and $\alpha$ spectroscopic factors are not as large as those
of the ground state, and the ratio of $S_\alpha$ to the ground state is
$S_{\alpha,\ell=0}(0^+_2)/S_{\alpha,\ell=2}(0^+_1)=0.56$. This reduction of the
$S_\alpha$ relative to the ground state reasonably  agrees with the observed values
which is in between 0.51 and 0.75~\cite{Peng1978,Berg1979,Tanabe1981}. It must be
emphasized that the $\alpha$ cluster correlations in the ground and $0^+_2$ states are
the origin of the large monopole transition strength (5.7 $\rm fm^2$) between these
states. In fact, if we exclude the \aSi molecular configurations from the GCM
calculation, the spectroscopic factors of the ground and $0^+_2$ states are reduced to
0.05 and 0.02 in the $\alpha$+${}^{28}{\rm Si}(0^+_1)$ channel, and the monopole
transition matrix is reduced to 2.32 $\rm fm^2$ which is smaller than the observed
value, 4.0 $\rm fm^2$.

The $0^+_3$ state has the largest overlap with the configuration shown in
Fig.~\ref{fig:density} (c) which amounts to 0.36. This state has similar magnitude of
the overlap with many other configurations on the $\beta$-constraint energy surface shown
in Fig.~\ref{fig:surface} (a), but it scarcely overlaps with the molecular configurations. 
Therefore, its RWA and spectroscopic factors are small, and we conclude that
the $0^+_3$ state is a $\beta$-vibration  state. This interpretation explains the large
monopole strength of this state, as it is well known that the $\beta$-vibration also
enhances the monopole transition strengths~\cite{Reiner1961}. Itoh
{\it et al.}~\cite{Itoh2013} observed three $0^+$ states (6.59, 7.65 and 7.95 MeV 
states) with the enhanced monopole strengths in this energy region, and 6.59 MeV state
plausibly coincides with the calculated $0^+_3$ state. However, neither of our
calculation nor other experiments reported additional $0^+$ states in between 6 to 8
MeV~\cite{Ouellet2011}. Therefore, more detailed  study is needed to confirm the 7.65
and 7.95 MeV states. 

The $0^+_4$ state is the superdeformed state which was already discussed in 
the previous AMD study~\cite{Kimura2004}. It has the large squared overlap (0.95) with
the configuration shown in Fig.~\ref{fig:density} (d). In addition, it also has
large overlap with the \OO configuration shown in Fig.~\ref{fig:density} (e), 
that amounts to 0.92. Hence, the superdeformed state of \Sal~is regarded as an 
\OO molecular state; {\it i.e.} it has a duality of the superdeformation
and clustering. From the observed strong monopole transitions, Itoh
{\it et al.} proposed the 10.49, 11.62 and 11.90 MeV states as the superdeformed
states. However, in contrast to their assignment, the present calculation shows that the
monopole transition  to the superdeformed state is negligible. This result clearly
reflects the nature of the monopole transition. As explained by 
Yamada {\it et al.}~\cite{Yamada2008}, the monopole transition excites the molecular
configurations which are contained in the ground state. In other words, the molecular
configurations orthogonal to the ground state at zero inter-cluster
distance are not populated by the monopole transition. Because the \OO configuration is
orthogonal to the ground state, the monopole transition from the ground state to the \OO
molecular state is strictly forbidden. Therefore, the present result does not
support the assignment of the \OO molecular state and the superdeformed state observed
in the $\alpha$ inelastic scattering experiment.

The $0^+_5$ and $0^+_6$ states are the highly excited \aSi molecular states which
overlap with both the S- and L-type \aSi configurations shown in
Figs.~\ref{fig:density} (f)-(i). As seen in Table~\ref{tab:mono}, these states 
are predominated by the $\alpha$+${}^{28}{\rm Si}(0^+_1)$ channel, while the ground
state and $0^+_2$ states are the mixture of the  $\alpha$+${}^{28}{\rm Si}(0^+_1)$ and
$\alpha$+${}^{28}{\rm Si}(2^+_1)$ channels. This is because of the weak interaction
between the clusters in the $0^+_5$ and $0^+_6$ states, which de-excites the
${}^{28}{\rm Si}$ cluster to its ground state (weak cluster polarization). Note that the
RWA of the $0^+_5$ and $0^+_6$ states have a peak at approximately 6 fm, which indicates
the large inter-cluster distance and enhanced clustering.
Owing to this pronounced \aSi molecular structure, these states have large
monopole transition strengths, and they may correspond to any of 
the 10.49, 11.62 and 11.90 MeV states observed by Itoh {\it et al.}. Interestingly, the
\aSi molecular states observed by L\"onnroth {\it et al.} are located at the same
energy region, and we consider that they are the same \aSi molecular states.

Thus, the present calculation has revealed the characteristics of the excited $0^+$
states. The monopole transition from the ground state has a selectivity, because the
ground state is a mixture of the deformed mean-field and \aSi molecular structure.
The $\beta$-vibration state and \aSi molecular states are strongly excited, but the \OO
molecular state (and hence the superdeformed state) is not. We conclude that many
of the states with enhanced monopole strengths observed below 12 MeV should be
attributed to the \aSi molecular states.  

\subsection{Assignment of the rotational bands}

\begin{figure}[tbp]
\includegraphics[width=0.6\hsize]{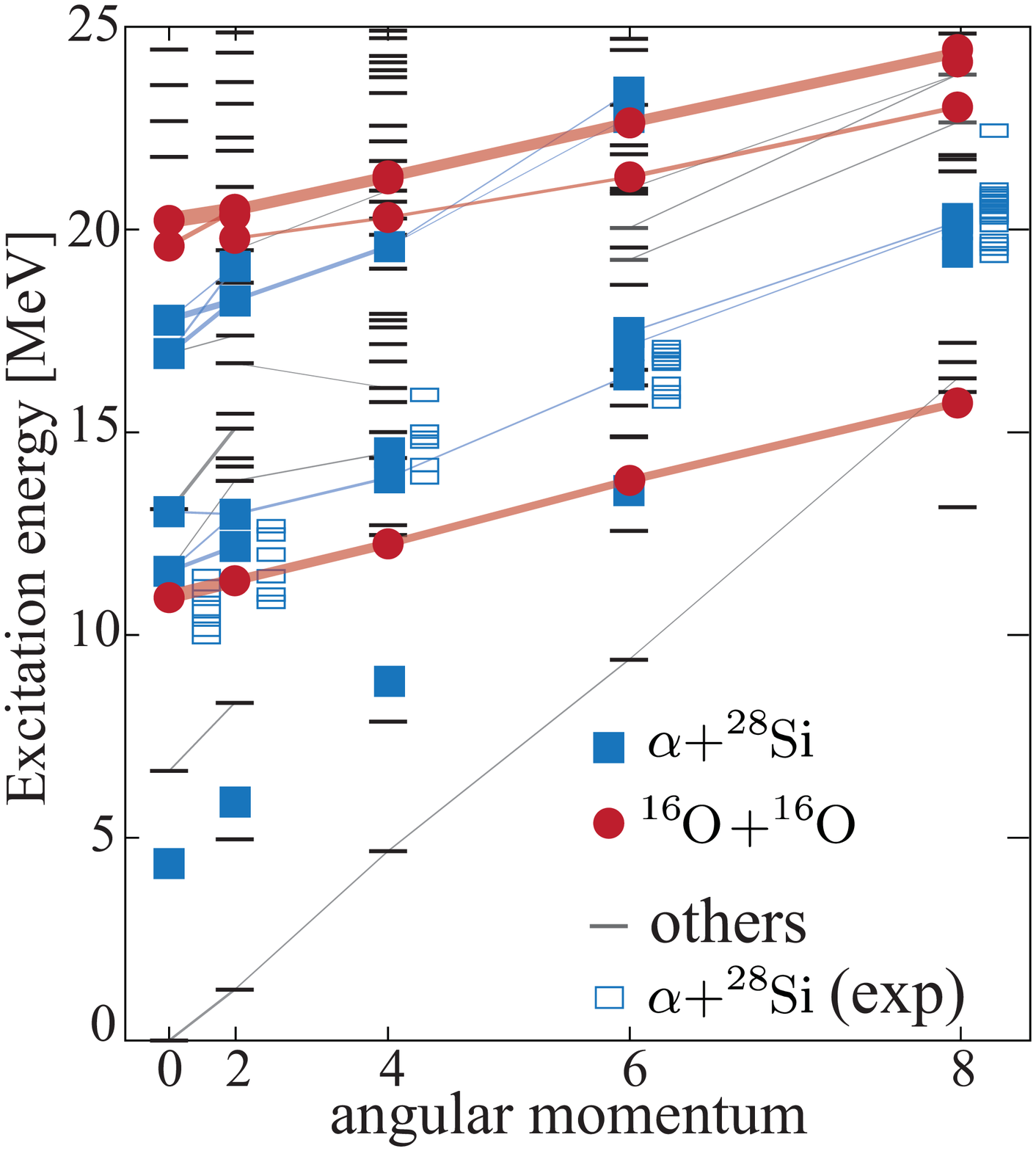}
 \caption{(color online) The calculated molecular bands up to $J^\pi=8^+$ states. The
 filled squares and  circles show the \aSi and \OO molecular states, respectively. The
 open squares show the  observed candidates of the \aSi molecular
 band~\cite{Lonnroth2010}. The $B(E2\uparrow)$  transitions stronger than 150 $e^2\rm
 fm^4$ are shown by the connecting lines whose widths are proportional to the
 magnitude of the transition matrices.} 
 \label{fig:be2}
\end{figure}

Figure~\ref{fig:be2} shows our assignment of the \aSi and \OO molecular bands from
the present calculation. The assignment is based on the calculated spectroscopic
factors. That is, if the spectroscopic factors in the \OO channel or if the sum of the
spectroscopic factors in the $\alpha$+${}^{28}{\rm Si}(0^+_1)$ and
$\alpha$+${}^{28}{\rm Si}(2^+_2)$ channels is larger than 0.10, we have assigned the
state as the molecular state. The figure also shows the $B(E2\uparrow)$ strengths larger
than 150 $e^2\rm fm^4$, which confirms that most of the molecular states
are connected by the strong $B(E2)$ transitions due to their strong quadrupole
deformation.

The assignment of the \OO band is essentially same with that proposed in the
previous AMD study and rather unique as it does not strongly fragment into many
states. The lowest \OO band is built on the $0^+_4$ state at 11.0 MeV, and as already
discussed in Ref.~\cite{Kimura2004}, it is identical to the superdeformed band with huge
moment-of-inertia as large as $\hbar^2/(2\mathcal I)=68$ keV.
Another \OO molecular band, in which the relative  motion between \Oxy~clusters is
excited, exists at approximately $E_x=20$ MeV, and the member states of this band with
$J\geq 2$ are fragmented into two or three states. 

The assignment of the \aSi band is not as unique as the \OO case since the member
states are fragmented into many states due to the strong coupling of the 
$\alpha$+${}^{28}{\rm Si}(0^+_1)$ and $\alpha$+${}^{28}{\rm Si}(2^+_2)$ channels, as
well as the coupling with the non-cluster configurations. There are many states which
have small but 
non-negligible spectroscopic factors in the \aSi~channels. For example, the states which
have the spectroscopic factors larger than 0.05 are almost twice as many as those shown in
Fig.~\ref{fig:be2}. This result is consistent with the observation by L\"onnroth
{\it et al.} who reported many excited states which have small fraction of the
\aSi~spectroscopic factors. However, for the sake of clarity and simplicity, here, we 
discuss the states with sufficiently large spectroscopic factors (larger than 0.10). We
suggest an \aSi band built on the $0^+_5$ and $0^+_6$ states. Although the member states
are considerably fragmented, it can be confirmed that many states are connected by the
strong $B(E2)$  transitions. We consider that this band corresponds to the \aSi band
reported by L\"onnroth {\it et al.} as the energies of the member states plausibly agree
with their observation. We also comment that the other band, in which the relative
motion of the clusters is excited, may be built on the $0^+$ states approximately at 17
MeV. We can see the candidates of the band member states up to the $J^\pi=6^+$, although
the  fragmentation is rather strong. Experimentally, several candidates of the \aSi
states have been reported above 15 MeV without firm spin-parity
assignment~\cite{Leachman1972,Obst1972}, and the present results may explain these 
observations.

\section{summary}\label{sec:summary}
We have investigated the properties of the \aSi and \OO molecular states in the
\Sal~excited states. An extended framework of AMD has been applied for handling both
of the \aSi and \OO channels in addition to the rotation effect of the deformed
\Si~cluster. It was found that the isoscalar monopole transition has the strong
selectivity to the molecular states: It strongly excites the \aSi molecular states, but
is inactive to the \OO molecular states. This selectivity originates in the different 
asymptotic behavior of the  molecular configurations at zero inter-cluster distance. 
We found that the assignment of the \aSi molecular states proposed by L\"onnroth
{\it et al.} reasonably agrees with the present calculation, while the \OO molecular
state or the superdeformed state proposed by Itoh {\it et al.} does not, as
the monopole transition strengths of the \OO molecular states are rather weak and is
not excited strongly by the $\alpha$ inelastic scattering.

\begin{acknowledgments}
This work was supported by a grant for the RCNP joint research project, the
collaborative research program 2020 at Hokkaido University, and JSPS KAKENHI Grant
No. 19K03859. 
\end{acknowledgments}

\bibliography{main}

\end{document}